\documentclass[a4paper]{scrartcl}
\pdfoutput=1
\usepackage[utf8]{inputenc}

\usepackage[T1]{fontenc} 
\usepackage{lmodern} 
\usepackage[sc,osf]{mathpazo} 
\usepackage{myT1classico} 
\DeclareMathAlphabet{\mathsf}{T1}
  {\sfdefault}{m}{n} 
\SetMathAlphabet{\mathsf}{bold}{T1}{\sfdefault}{b}{n} 

\usepackage{amsmath,amssymb,amsthm}
\usepackage{tensor}
\usepackage{accents}

\usepackage{enumitem}

\usepackage{relsize}
\newcommand{\acronym}[1]{\texorpdfstring{\textsmaller{#1}}{#1}}

\usepackage[british]{babel}

\usepackage{microtype} 

\usepackage{authblk}

\usepackage{todonotes}

\usepackage[colorlinks]{hyperref}
\usepackage[nameinlink]{cleveref}

\usepackage[style=phys,biblabel=brackets,eprint=true,sorting=noney,
  sortcites=false]{biblatex}
\newbibmacro{string+doiurl}[1]{%
  \iffieldundef{doi}{%
    \iffieldundef{url}{#1}{\href{\thefield{url}}{#1}}%
  }{\href{https://doi.org/\thefield{doi}}{#1}}%
}
\DeclareFieldFormat{title}{\usebibmacro{string+doiurl}{\mkbibemph{#1}}}
\DeclareSortingTemplate{noney}{
  \sort{\citeorder}
}

\newcommand{\e}{\mathrm{e}} 
\newcommand{\D}{\mathrm{d}} 
\newcommand{\DD}{\mathrm{D}} 
\newcommand{\R}{\mathbb R}

\newcommand{\el}{\accentset{\lambda}{e}}
\newcommand{\nablal}{\accentset{\lambda}{\nabla}}

\theoremstyle{plain}
\newtheorem{theorem}{Theorem}
\newtheorem{proposition}[theorem]{Proposition}
\newtheorem{claim}[theorem]{Claim}

\title{Comment on `The Classical Limit of Teleparallel Gravity'}

\author[1,a]{Philip K. Schwartz}
\author[1,2,b]{Arian L. von Blanckenburg}
\affil[1]{Institute for Theoretical Physics,
  Leibniz University Hannover, \par
  Appelstraße 2, 30167 Hannover, Germany}
\affil[2]{Max Planck Institute for Gravitational Physics (Albert
  Einstein Institute), \par
  Callinstraße 38, 30167 Hannover, Germany}
\affil[a]{\normalfont\texttt{\href{mailto:philip.schwartz@itp.uni-hannover.de}
    {philip.schwartz@itp.uni-hannover.de}}}
\affil[b]{\normalfont\texttt{\href{mailto:arian.von.blanckenburg@aei.mpg.de}
    {arian.von.blanckenburg@aei.mpg.de}}}

\date{}

\begin{document}
\maketitle

\begin{abstract}
  \noindent
  We critically discuss the claims of a recent article regarding the
  Newtonian limit of the teleparallel equivalent of general relativity
  (\acronym{TEGR}) in pure-tetrad formulation \cite{Meskhidze:2024}.
  In particular, we refute this article's purported main result that
  if a regular Newtonian limit exists, the torsion of the limiting
  derivative operator (i.e., connection) necessarily vanishes.
\end{abstract}

\section{Introduction}
\label{sec:intro}

In the recent article \cite{Meskhidze:2024}, it is claimed that when
geometrically formulating the Newtonian limit of the teleparallel
equivalent of general relativity (\acronym{TEGR}) in so-called
pure-tetrad formulation---i.e.\ in the formulation of \acronym{TEGR}
in the \emph{Weitzenböck gauge}, working with an orthonormal frame and
the connection / derivative operator parallelising the frame---in the
style of Newton--Cartan gravity, the torsion vanishes in the Newtonian
limit.  In this comment, we provide a counterexample and try to
pinpoint the oversight in the attempted proof of this claim in ref.\
\cite{Meskhidze:2024}.  In addition, we critically discuss some of the
interpretational considerations made in ref.\ \cite{Meskhidze:2024},
and give some remarks in hope to clarify possible confusion regarding
the term `torsional Newton--Cartan geometry'.
\pagebreak

\section{Claimed main result of ref.\ \cite{Meskhidze:2024}}

The setting and terminology of reference \cite{Meskhidze:2024} is as
follows.
\begin{itemize}
\item A \emph{classical spacetime model} is a four-dimensional
  differentiable manifold $M$ with a nowhere-vanishing one-form $t_a$
  and a symmetric two-tensor field $h^{ab}$ of signature $(0,1,1,1)$
  such that $t_a h^{ab} = 0$, as well as a covariant derivative
  operator $\nabla_a$ compatible with $t_a$ and $h^{ab}$.  Mostly,
  only the case that $t_a$ be closed is considered.
\item A \emph{classical tetrad} for the metric structure given by
  $t_a$ and $h^{ab}$ is a local frame $\{(f_i)^a\}$ ($i = 1, \dots,
  4$) of vector fields on $M$ such that for the cotetrad / dual frame
  $\{(f^i)_a\}$, we have $(f^1)_a = t_a$ and $(f^i)_a (f^j)_b h^{ab} =
  \delta^{ij}$ for $i,j \in \{2,3,4\}$.  Starting just with the tetrad
  and its dual, one can recover the `metrics' as $t_a = (f^1)_a$ and
  $h^{ab} = \sum_{i=2}^4 (f_i)^a (f_i)^b$.
\item Given a Lorentzian metric $g_{ab}$ , we can consider local
  orthonormal frames $\{(e_i)^a\}$ ($i = 1, \dots, 4$), satisfying
  $g_{ab} (e_i)^a (e_j)^b = \eta_{ij}$ where $\eta_{ij}$ are the
  components of the Minkowski metric.  We can write the metric in
  terms of the cotetrad $\{(e^i)_a\}$ as $g_{ab} = \sum_{i,j = 1}^4
  \eta_{ij} (e^i)_a (e^j)_b = \sum_{i = 1}^4 \eta_{ii} (e^i)_a
  (e^i)_b$, and the inverse metric in terms of the tetrad as $g^{ab} =
  \sum_{i,j = 1}^4 \eta^{ij} (e_i)^a (e_j)^b = \sum_{i = 1}^4
  \eta^{ii} (e_i)^a (e_i)^b$.
\item For a tetrad $\{(\el_i)^a\}$ depending on a parameter $\lambda >
  0$ (to be interpreted as a Lorentzian tetrad for a
  $\lambda$-dependent metric, with the limit $\lambda \to 0$
  considered the `classical', i.e.\ Newtonian, limit), the following
  limiting conditions are introduced:\footnote{Here we have corrected
    a small error in the version of \ref{item:conv_1} from ref.\
    \cite{Meskhidze:2024}, based on the false claim that for a
    classical cotetrad only the first `leg' $(f^1)_a$ is
    non-vanishing.  Of course, since the cotetrad is pointwise a basis
    of the cotangent space, none of its elements can vanish anywhere
    on its domain.  This error however does not have any
    consequences.}
  \begin{enumerate}[label=C\arabic**]
  \item \label{item:conv_1} $\sum_{i=1}^4 (\el^i)_a \to (f^1)_a = t_a$
    as $\lambda \to 0$ for some closed field $t_a$, and
  \item \label{item:conv_2} $\sqrt{\lambda} \sum_{i=1}^4 (\el_i)^a \to
    \sum_{i=2}^4 (f_i)^a$ as $\lambda \to 0$.
  \end{enumerate}
\end{itemize}

The main claim of ref.\ \cite{Meskhidze:2024} is now the following
(where we have added the definition of $\nablal$ to make the statement
self-contained):
\begin{claim}[Theorem 1 of ref.\ \cite{Meskhidze:2024}]
  \label{claim}
  Suppose that $\{(\el_i)^a\}$ is a one-parameter family of tetrads on
  a manifold $M$.  Suppose $\{(f_i)^a\}$ satisfies conditions
  \ref{item:conv_1} and \ref{item:conv_2}.  For each value of
  $\lambda$, let $\nablal_a$ be the unique (flat) derivative operator
  on $M$ for which the tetrad fields $(\el_i)^a$ are parallel, i.e.\
  $\nablal_a (\el_i)^b = 0$.  Suppose that there is a derivative
  operator $\nabla_a$ on $M$ satisfying $\nablal_a \to \nabla_a$ as
  $\lambda \to 0$.  Then, the following holds.
  \begin{enumerate}
  \item $\nabla_a$ is such that $(M, \{(f_i)^a\}, \nabla_a)$ is a
    classical spacetime model where $\{(f_i)^a\}$ is a classical
    tetrad and $\nabla_a$ is flat.
  \item \label{claim_2} For any derivative operator $\nabla_a$
    satisfying the above, the torsion vanishes.
  \end{enumerate}
\end{claim}
The important part of the claimed result is the second one: if
correct, it would mean that in the Newtonian limit of \acronym{TEGR}
the torsion of the limiting derivative operator vanishes.

However, this claim turns out to be wrong: we are going to construct
explicit counterexamples to the torsion vanishing in the $\lambda \to
0$ limit, and also point out the mistake in the proof attempt in
ref.\ \cite{Meskhidze:2024}.

\section{Infinite counterexamples}
\label{sec:counterexamples}

We can obtain counterexamples to part \ref{claim_2} of \cref{claim} as
follows.
\begin{proposition}
  \label{prop:counterex}
  Let $M$ be a four-dimensional manifold and $\{(f_i)^a\}$, $i = 1,
  \dots, 4$, a tetrad / frame of vector fields for it such that
  \begin{enumerate}[label=(\roman*)]
  \item the first `leg' $(f^1)_a$ of the dual frame is closed, and
  \item the tetrad is anholonomic---i.e., the fields $(f_i)^a$ do not
    all commute everywhere, i.e.\ some of their Lie brackets
    \begin{equation}
      [\boldsymbol{f}_i, \boldsymbol{f}_j]^a
      = (f_i)^b \partial_b (f_j)^a - (f_j)^b \partial_b (f_i)^a
    \end{equation}
    are non-zero at some points.  (Here, we use boldface symbols for
    fields when not using abstract index notation.)
  \end{enumerate}
  We set $(\el_1)^a := (f_1)^a$ and $(\el_i)^a :=
  \frac{1}{\sqrt{\lambda}} (f_i)^a$ for $i = 2, 3, 4$.  Then the
  manifold $M$ together with the frames $\{(\el_i)^a\}$ and
  $\{(f_i)^a\}$ satisfies the assumptions of \cref{claim}, but the
  limiting derivative operator $\nabla_a$ has non-vanishing torsion.

  \begin{proof}
    The dual frame is given by $(\el^1)_a = (f^1)_a$ and $(\el^i)_a =
    \sqrt{\lambda} (f^i)_a$ for $i = 2, 3, 4$.  Thus, by direct
    computation we see that conditions \ref{item:conv_1} and
    \ref{item:conv_2} are satisfied.

    Further, we denote by $\nabla_a$ the (unique) derivative operator
    with respect to which the frame $\{(f_i)^a\}$ is parallel.  By
    construction, for all values of $\lambda$ it also satisfies
    $\nabla_a (\el_1)^b = \nabla_a (f_1)^b = 0$ and, for $i = 2,3,4$,
    $\nabla_a (\el_i)^b = \nabla_a \big(\tfrac{1}{\sqrt{\lambda}}
    (f_i)^b \big) = \tfrac{1}{\sqrt{\lambda}} \nabla_a(f_i)^b = 0$.
    This shows that the family $\nablal_a$ of derivative operators
    defined by $\nablal_a (\el_i)^b = 0$ is actually independent of
    $\lambda$: we have $\nablal_a = \nabla_a$.\footnote{This may also
      be seen by direct computation of the connecting field of
      $\nablal_a$ relative to a coordinate derivative operator
      $\partial_a$, which is given by
      \begin{equation*}
        \tensor{\accentset{\lambda}{C}}{^a_{bc}}
        = \sum_{i=1}^4 (\el^i)_c \partial_b(\el_i)^a
        = (f^1)_c \partial_b(f_1)^a
          + \sum_{i=2}^4 \sqrt{\lambda} (f^i)_c \partial_b
            \big(\tfrac{1}{\sqrt{\lambda}}(f_i)^a\big)
        = \sum_{i=1}^4 (f^i)_c \partial_b(f_i)^a \; .
      \end{equation*}}

    Therefore, the limiting derivative operator $\lim_{\lambda\to0}
    \nablal_a$ exists, and it is simply the unique derivative operator
    $\nabla_a$ for which the frame $\{(f_i)^a\}$ is parallel.  Since
    the frame is anholonomic, the torsion of this derivative operator
    is non-vanishing (explicitly, we have $\boldsymbol{T}
    (\boldsymbol{f}_i, \boldsymbol{f}_j) = [\boldsymbol{f}_i,
    \boldsymbol{f}_j]$).
  \end{proof}
\end{proposition}

Tetrads satisfying the conditions of this counterexample proposition
can easily be obtained by combining an anholonomic triad on a
three-dimensional manifold with an independent `time direction':
\begin{proposition}
  Let $N$ be a three-dimensional manifold and $\{(E_i)^A\}$, $i = 2,
  3, 4$, an anholonomic frame of vector fields for $N$.

  We consider $M = \R \times N = \{(t, x^A)\}$.  On $M$, we define the
  vector fields $(f_1)^a = (\frac{\partial}{\partial t})^a$ and
  \begin{equation}
    (f_i)^a = (E_i)^A (\tfrac{\partial}{\partial x^A})^a
    \; \text{for} \; i = 2, 3, 4:
  \end{equation}
  the fields $(f_i)^a$ for $i = 2, 3, 4$ are the fields $(E_i)^A$
  `understood as $t$-independent vector fields on $M = \R \times N$'.
  Put differently, they are defined by the condition $\DD
  \pi|_{(t,x^A)} (\boldsymbol{f}_j|_{(t,x^A)}) = \boldsymbol{E}_j
  |_{(x^A)}$, where $\pi \colon \R \times N \to N$ is the projection
  onto the second factor.  By construction, $\{(f_i)^a\}$, $i = 1,
  \dots, 4$ is a frame for $M$ satisfying the assumptions of
  \cref{prop:counterex} (the first `leg' of the dual frame is
  $\boldsymbol{f}^1 = \boldsymbol{\D}t$, which is exact and therefore
  closed).  Thus we obtain a counterexample to part \ref{claim_2} of
  \cref{claim}.  \qed
\end{proposition}

Most concretely, we can take $\{(E_i)^A\}$ to be an anholonomic frame
for $N = \R^3$.  This provides an infinite family of counterexamples.
For example, writing $\R^3 = \{(x,y,z)\}$, we can define
$\boldsymbol{E}_2 = \cos(x) \frac{\partial}{\partial x} + \sin(x)
\frac{\partial}{\partial y}$, $\boldsymbol{E}_3 = -\sin(x)
\frac{\partial}{\partial x} + \cos(x) \frac{\partial}{\partial y}$,
$\boldsymbol{E}_4 = \frac{\partial}{\partial z}$, obtaining the
non-vanishing Lie bracket $[\boldsymbol{E}_2, \boldsymbol{E}_3] = -
\frac{\partial}{\partial x}$.

\section{The attempted proof in ref.\ \cite{Meskhidze:2024}}

The strategy of the proof attempt in ref.\ \cite{Meskhidze:2024} of
part \ref{claim_2} of \cref{claim} (i.e.\ that the torsion necessarily
vanishes in the Newtonian limit) is as follows.  Since for each value
of $\lambda$ the derivative operator $\nablal_a$ in question is
defined by the tetrad $\{(\el_i)^a\}$ being parallel, it is compatible
with the metric $g_{ab}(\lambda)$ defined by the tetrad
$\{(\el_i)^a\}$.  Hence, the connecting field
$\tensor{C}{^a_{bc}}(\lambda)$ of $\nablal_a$ relative to a coordinate
derivative operator $\partial_a$ may be expressed in terms of its
torsion and the metric by the Koszul formula,
\begin{align}
  \label{eq:connecting_field_koszul}
  \tensor{C}{^a_{bc}}(\lambda)
  = \frac{1}{2} g^{ad}(\lambda)
  \Big[ &\partial_d g_{bc}(\lambda) - \partial_b g_{dc}(\lambda)
          - \partial_c g_{db}(\lambda) \nonumber \\
        &- g_{dm}(\lambda) \tensor{T}{^m_{cb}}(\lambda)
          + g_{bm}(\lambda) \tensor{T}{^m_{dc}}(\lambda)
          + g_{cm}(\lambda) \tensor{T}{^m_{db}}(\lambda) \Big] \, .
\end{align}
Since $\nablal_a$ is defined by $\{(\el_i)^a\}$ being parallel, its
torsion may be expressed in terms of the tetrad as
\begin{equation}
  \tensor{T}{^a_{bc}}(\lambda)
  = \sum_{i=1}^4 (\el^i)_{[c} \partial_{b]} (\el_i)^a \; .
\end{equation}
Then ref.\ \cite{Meskhidze:2024} uses this expression for the torsion
and the expressions
\begin{subequations}
\begin{align}
  g_{ab}(\lambda)
  &= \sum_{i,j=1}^4 \eta_{ij} (\el^i)_a (\el^j)_b
  = \sum_{i=1}^4 \eta_{ii} (\el^i)_a (\el^i)_b \; , \\
  g^{ab}(\lambda)
  &= \sum_{i,j=1}^4 \eta^{ij} (\el_i)^a (\el_j)^b
  = \sum_{i=1}^4 \eta^{ii} (\el_i)^a (\el_i)^b
\end{align}
\end{subequations}
for the metric and its inverse to explicitly compute all terms in the
Koszul formula \eqref{eq:connecting_field_koszul} in terms of the
tetrad.  Using the assumed limiting behaviour of the tetrads as
$\lambda \to 0$, ref.~\cite{Meskhidze:2024} then attempts to show that
in general the torsion terms contain negative powers of $\lambda$,
such that the connecting field would not converge in the limit
$\lambda \to 0$.  If the derivative operator converges, the
corresponding terms hence would need to vanish, which is shown to
imply that the limiting derivative operator $\nabla_a$ would need to
be torsion-free.  However, the computation of the torsion terms
contains some crucial oversights, such that the just described proof
strategy of ref.\ \cite{Meskhidze:2024} breaks down.

The basic type of mistake in ref.\ \cite{Meskhidze:2024} in the
calculation of the limit of the connecting field
\eqref{eq:connecting_field} is that in general the sum of products of
individual factors is not the product of the sums of the factors,
i.e.\ that
\begin{equation}
  \label{eq:error_sum_prod}
  \sum_{k=1}^n A_k B_k
  \ne \left(\sum_{k=1}^n A_k\right) \left(\sum_{l=1}^n B_l\right).
\end{equation}
At several stages in the attempted proof of \cref{claim} in ref.\
\cite{Meskhidze:2024}, a sum of products is mistakenly replaced by the
corresponding product of sums.  The first time this happens in the
proof is at the bottom of page 15 of ref.\ \cite{Meskhidze:2024} in
the computation of the explicit form of the inverse metric.  The
leading-order term (of order $1/\lambda$) of the result here is
however correct (since a replacement as in \eqref{eq:error_sum_prod}
is made twice), such that this does not affect the conclusion
regarding the torsion.\footnote{Note that the incorrect higher-order
  terms nevertheless lead to missing terms in the order-$\lambda^0$
  part of the explicit form for the connecting field presented later
  in ref.\ \cite{Meskhidze:2024}.}  Similarly, in equation (4.2) of
ref.\ \cite{Meskhidze:2024} a `sum of products'-mistake is made,
affecting however only higher-order terms.

The point in ref.\ \cite{Meskhidze:2024} where the consequences of
replacing a sum of products by a product of sums become significant is
in the calculation of the first torsion term from the Koszul formula,
in equation (4.3).  In this equation, the replacement leads to a
$1/\sqrt{\lambda}$ term, which in the end is responsible for the
conclusion that the limit of the derivative operator can only exist if
the torsion vanishes: due to the incorrect replacement, in (4.3) of
ref.\ \cite{Meskhidze:2024} the $\lambda$-expanded form of $(\el^1)_b$
is multiplied with a derivative of (the sum of) $(\el_j)^m$, which has
a leading term of order $1/\sqrt{\lambda}$; thus leading to the
divergent term.  However, evaluating the sum of products correctly, we
encounter the derivatives $\partial_c (\el_j)^m$ only when multiplied
with $(\el^j)_b$ (and summed over $j$).  Therefore, the
$1/\sqrt{\lambda}$ from the leading term of $\partial_c (\el_j)^m$
will get cancelled by $\sqrt{\lambda}$ from the leading term of
$(\el^j)_b$ (see our discussion of the limiting assumptions on the
tetrad below), and no term arises that would diverge in the $\lambda
\to 0$ limit.

This shows explicitly how in ref.\ \cite{Meskhidze:2024} the
replacement of a sum of products by a product of sums leads to the
wrong conclusion of torsion necessarily vanishing in the limit, i.e.\
where the argumentation in the attempted proof of part \ref{claim_2}
of \cref{claim} goes wrong.

\section{The limiting assumptions}
\label{sec:limit_assump}

The oversight explained above of replacing a sum of products by a
product of sums \eqref{eq:error_sum_prod} does not only affect the
attempted proof of \cref{claim} in ref.\ \cite{Meskhidze:2024}, but
even the assumptions \ref{item:conv_1} and \ref{item:conv_2} on the
limiting behaviour of the tetrad themselves: only due to the erroneous
replacement of sums of products by products of sums, the proof attempt
in ref.\ \cite{Meskhidze:2024} encounters the summed expressions
$\sum_{i=1}^4 (\el^i)_a$ and $\sum_{i=1}^4 (\el_i)^a$ on whose limits
\ref{item:conv_1} and \ref{item:conv_2} make assumptions.  To
correctly evaluate the sums in the calculation, one needs knowledge
about the behaviour of the vector fields $(\el_i)^a$ and the covector
fields $(\el^i)_a$ \emph{themselves}, not just their sums.

Assumptions \ref{item:conv_1} and \ref{item:conv_2} are introduced in
ref.\ \cite{Meskhidze:2024} in order to propose analogues in the
tetrad case of the limiting assumptions on the Lorentzian metric and
its inverse as discussed by Malament for the Newtonian limit of
standard \acronym{GR} \cite{Malament:1986}.  As quoted in
ref.~\cite{Meskhidze:2024}, the latter read as follows:
\begin{enumerate}[label=C\arabic*]
\item \label{item:metric_conv_1} $g_{ab}(\lambda) \to t_a t_b$ as
  $\lambda \to 0$ for some closed field $t_a$, and
\item \label{item:metric_conv_2} $\lambda g^{ab}(\lambda) \to -h^{ab}$
  as $\lambda \to 0$ for some field $h^{ab}$ of signature $(0,1,1,1)$.
\end{enumerate}
Expressing the Lorentzian (inverse) metric and the classical `metrics'
in terms of a Lorentzian and a classical (co-)tetrad, these conditions
take the form
\begin{subequations} \label{eq:metric_conv}
\begin{align}
  \label{eq:metric_conv_1}
  \sum_{i,j = 1}^4 \eta_{ij} (\el^i)_a (\el^j)_b
  &\to (f^1)_a (f^1)_b \; , \\
  \label{eq:metric_conv_2}
  \lambda \sum_{i,j = 1}^4 \eta^{ij} (\el_i)^a (\el_j)^b
  &\to -\sum_{i=2}^4 (f_i)^a (f_i)^b \; .
\end{align}
\end{subequations}
Comparing with the proposed assumptions \ref{item:conv_1} and
\ref{item:conv_2} on the tetrad, one realises that these are quite
different in nature: while \ref{item:metric_conv_1} and
\ref{item:metric_conv_2} in form of \eqref{eq:metric_conv} are
assumptions on sums of products, \ref{item:conv_1} and
\ref{item:conv_2} are assumptions on the corresponding sums of the
individual factors.  The attempt to replace \eqref{eq:metric_conv} by
\ref{item:conv_1} and \ref{item:conv_2} thus bears some resemblance to
the earlier-described incorrect equating a sum of products with a
product of sums.  In fact, assumptions \ref{item:conv_1} and
\ref{item:conv_2} seem insufficient (in the mathematical sense) as a
replacement of \ref{item:metric_conv_1} and \ref{item:metric_conv_2}
in the tetrad case: one can easily construct explicit examples showing
that they are weaker than \eqref{eq:metric_conv}.\footnote{They are
  `weaker' in the sense that there are one-parameter families of
  tetrads satisfying \ref{item:conv_1} and \ref{item:conv_2}, but not
  \eqref{eq:metric_conv}.  Explicitly, fixing a classical tetrad
  $\{(f_i)^a\}$, we can define $(\el_1)^a := (f_1)^a$, $(\el_2)^a :=
  \frac{1}{\sqrt{\lambda}} (f_2)^a$, $(\el_3)^a :=
  \frac{1}{\sqrt{\lambda}} (\frac{1}{\lambda} (f_3)^a + (f_3)^a)$,
  $(\el_4)^a := \frac{1}{\sqrt{\lambda}} (-\frac{1}{\lambda} (f_3)^a +
  (f_4)^a)$.  The dual tetrad is then given by $(\el^1)_a = (f^1)_a =
  t_a$, $(\el^2)_a = \sqrt{\lambda} (f^2)_a$, $(\el^3)_a =
  \sqrt{\lambda} (\frac{\lambda}{\lambda+1} (f^3)_a + \frac{1}{\lambda
    + 1} (f^4)_a)$, $(\el^4)_a = \sqrt{\lambda} (f^4)_a$.  A direct
  computation shows that these satisfy \ref{item:conv_1},
  \ref{item:conv_2}, and \eqref{eq:metric_conv_1}, but fail to satisfy
  \eqref{eq:metric_conv_2}: the rescaled inverse metric $\lambda
  g^{ab}(\lambda) = \lambda \sum_{i,j = 1}^4 \eta^{ij} (\el_i)^a
  (\el_j)^b$ even fails to converge.}

Combining our above observations, one might ask for a set of
assumptions that is
\begin{enumerate}[label=(\alph*),nosep]
\item formulated in terms of the one-parameter families of tetrad
  vector fields $(\el_i)^a$ and the dual covector fields $(\el^i)_a$
  themselves (instead of specific sums thereof), and
\item sufficient for \eqref{eq:metric_conv}, such that it can act as a
  proper replacement for \ref{item:metric_conv_1} and
  \ref{item:metric_conv_2} in the tetrad case.
\end{enumerate}
A natural guess for such a set of assumptions, motivated from the
metric convergence conditions \eqref{eq:metric_conv}, is given by
\begin{subequations}
\label{eq:new_conv}
\begin{align}
  \label{eq:new_conv_dual_frame}
  (\el^1)_a
  &\xrightarrow{\lambda\to0} (f^1)_a = t_a \; ,
  &\frac{1}{\sqrt{\lambda}} (\el^i)_a
  &\xrightarrow{\lambda\to0} (f^i)_a \; \text{for} \; i = 2,3,4,\\
  \label{eq:new_conv_frame}
  (\el_1)^a
  &\xrightarrow{\lambda\to0} (f_1)^a \; ,
  &\sqrt{\lambda} (\el_i)^a
  &\xrightarrow{\lambda\to0} (f_i)^a \; \text{for} \; i = 2,3,4.
\end{align}
\end{subequations}
Note that it is sufficient to require either conditions
\eqref{eq:new_conv_dual_frame} on the cotetrad or
\eqref{eq:new_conv_frame} on the tetrad: either combination implies
the other.  (The same is not true when `mixing' them: the combination
of the first condition of \eqref{eq:new_conv_dual_frame} with the
second of \eqref{eq:new_conv_frame}, or vice versa, is weaker than all
of \eqref{eq:new_conv}.)  In fact, in ref.\ \cite{Meskhidze:2024} it
is actually stated that the desired limiting behaviour of the first
(i.e.\ timelike) `leg' of the cotetrad is according to the first
condition of \eqref{eq:new_conv_dual_frame}, and that of the spacelike
fields of the tetrad according to the second condition of
\eqref{eq:new_conv_frame}, respectively.  However, immediately after
stating this, ref.\ \cite{Meskhidze:2024} attempts to `formalize'
these behaviours in the assumptions \ref{item:conv_1} and
\ref{item:conv_2}---which are too weak to allow conclusions on the
metric's behaviour, as discussed above.

In fact, the convergence assumptions \eqref{eq:new_conv} on the tetrad
and/or dual tetrad are not only naturally \emph{motivated} from the
metric convergence conditions \ref{item:metric_conv_1} and
\ref{item:metric_conv_2}, but they even essentially \emph{follow} from
the latter: as we (the present authors) show in the accompanying
article \cite{Schwartz.vonBlanck:2024b}, whenever one is given a
one-parameter family of Lorentzian metrics $g_{ab}(\lambda)$
converging to the metric structure $t_a, h^{ab}$ of a classical
spacetime model according to \ref{item:metric_conv_1} and
\ref{item:metric_conv_2}, any family of orthonormal frames for these
metrics converges pointwise to a classical tetrad according to
\eqref{eq:new_conv}, assuming that two `obvious' necessary conditions
are satisfied: the spatial frame must not rotate indefinitely as the
limit is approached, and the frame's boost velocity with respect to
some fixed reference observer needs to converge.\footnote{In reference
  \cite{Schwartz.vonBlanck:2024b}, we use different conventions for
  the signature and normalisation of Lorentzian metrics, namely
  $g_{ab}(\text{this comment}) = -\lambda g_{ab}(\text{reference
    \cite{Schwartz.vonBlanck:2024b}})$.  This also leads to a
  different distribution of factors of $\sqrt{\lambda}$ in the
  Lorentzian (co-)tetrads.}  Note however that our result in reference
\cite{Schwartz.vonBlanck:2024b} regards only pointwise convergence,
such that no conclusion on the smoothness of the limiting classical
tetrad can be made (except, of course, the smoothness of $(f^1)_a =
t_a$).  Smoothness of $\{(f_i)^a\}$, if desired, has to be added as an
extra assumption.

Note that our counterexamples constructed in
\cref{sec:counterexamples} satisfy conditions \eqref{eq:new_conv}.
This shows that the conclusion of part \ref{claim_2} of \cref{claim}
is still false when assumptions \ref{item:conv_1} and
\ref{item:conv_2} are replaced by the better-motivated (and stronger)
assumptions \eqref{eq:new_conv}.  Actually, one can quite easily
compute the limiting derivative operator, and hence its torsion, as
follows.  As stated in ref.\ \cite{Meskhidze:2024}, the connecting
field of the unique flat derivative operator for which the tetrad
$\{(\el_i)^a\}$ is parallel, relative to a coordinate derivative
operator $\partial_a$, is given by
\begin{equation}
  \label{eq:connecting_field}
  \tensor{\accentset{\lambda}{C}}{^a_{bc}}
  = \sum_{i=1}^4 (\el^i)_c \partial_b (\el_i)^a \, .
\end{equation}
We can use this formula for the calculation of the limit $\lambda \to
0$, which avoids the use of the much more complicated Koszul formula
\eqref{eq:connecting_field_koszul}.  Assumptions \eqref{eq:new_conv}
mean that, as $\lambda \to 0$, we have
\begin{subequations}
\begin{align}
  (\el^1)_a &= (f^1)_a + o(1), \\
  \frac{1}{\sqrt{\lambda}} (\el^i)_a
            &= (f^i)_a + o(1), \; i = 2, 3, 4, \\
  (\el_1)^a &= (f_1)^a + o(1), \\
  \sqrt{\lambda}(\el_i)^a &= (f_i)^a + o(1), \; i = 2, 3, 4.
\end{align}
\end{subequations}
Thus, for $\lambda$ small enough, the connecting field 
\eqref{eq:connecting_field} is given by
\begin{align}
  \tensor{\accentset{\lambda}{C}}{^a_{bc}}
  &= (\el^1)_c \partial_b (\el_1)^a
    + \sum_{i=2}^4 (\el^i)_c \partial_b (\el_i)^a \nonumber\\
  &= \left( (f^1)_c + o(1) \right) \partial_b \left( (f_1)^a + o(1)
    \right)
    + \sum_{i=2}^4 \sqrt{\lambda} \left( (f^i)_a + o(1) \right)
      \partial_b \frac{1}{\sqrt{\lambda}} \left( (f_i)^a + o(1)
    \right) \nonumber\\
  &= (f^1)_c \partial_b (f_1)^a + o(1)
    + \sum_{i=2}^4 \left( (f^i)_c \partial_b (f_i)^a + o(1) \right)
    \nonumber\\
  &= \sum_{i=1}^4 (f^i)_c \partial_b (f_i)^a + o(1),
\end{align}
which has a perfectly finite limit as $\lambda \to 0$, without the
need for any further assumptions: the limiting derivative operator
$\nabla_a$ has connecting field
\begin{subequations}
\begin{align}
  \tensor{C}{^a_{bc}}
  &= \sum_{i=1}^4 (f^i)_c \partial_b (f_i)^a,\\
  \intertext{i.e.\ it is the unique derivative operator for which the
  limiting classical tetrad $\{(f_i)^a\}$ is parallel.  In particular,
  its torsion is non-vanishing if (and only if) the classical tetrad
  $\{(f_i)^a\}$ is anholonomic.  Explicitly, it is given by}
  \tensor{T}{^a_{bc}}
  &= 2 \tensor{C}{^a_{[bc]}}
    = 2 \sum_{i=1}^4 (f^i)_{[c} \partial_{b]} (f_i)^a.
\end{align}
\end{subequations}

\section{Discussion}

The above discussion has shown in detail that the main claim of ref.\
\cite{Meskhidze:2024} is false: taking the Newtonian limit of
(pure-tetrad) teleparallel gravity, the torsion does \emph{not}
necessarily vanish in the limit.  Hence, all further points of ref.\
\cite{Meskhidze:2024} based on the vanishing of the torsion are,
unfortunately, making an incorrect assumption, and thus are void.  In
particular, it is not true that in the Newtonian limit of teleparallel
gravity, there are no gravitational effects.  On the contrary, the
limiting theory is in fact the one discussed by one of the present
authors in article \cite{Schwartz:2023}, which is equivalent to
standard Newton--Cartan gravity, and thus encompasses the full range
of standard Newtonian gravity---albeit in a specific geometric
formulation.  The equivalence between standard general relativity and
its teleparallel equivalent \emph{does} indeed survive the transition
to the Newtonian limit.  Therefore, consideration of the Newtonian
limit does in fact \emph{support} the view that there is
underdetermination regarding spacetime structure between standard
general relativity and its teleparallel equivalent (contrary to the
perspective taken in ref.\ \cite{Meskhidze:2024}, based on its alleged
findings on vanishing torsion).

We also want to comment on the role played by the so-called \emph{mass
  torsion} in the teleparallel versions of Newton--Cartan gravity
discussed in the above-mentioned article~\cite{Schwartz:2023} on the
Newtonian limit of \acronym{TEGR} and the article \cite{Read.Teh:2018}
on its null reduction.  These articles use an extension of Galilei
coframes (i.e., in the language of ref.\ \cite{Meskhidze:2024}, of
classical cotetrads) to so-called \emph{extended coframes}---these
include an extra one-form related to the `mass direction' of the
Bargmann algebra (the centrally extended Galilei algebra).  Using this
extension, one can define the notion of \emph{extended torsion} of a
derivative operator compatible with the classical spacetime structure,
which on top of the usual torsion includes an additional component
termed the mass torsion.  In ref.\ \cite{Meskhidze:2024}, it is
claimed that `both proposals [i.e.\ \cite{Schwartz:2023,
  Read.Teh:2018}] require generalizing the notion of torsion with the
extended vielbein formalism', since `the results in the present paper
[i.e.\ \cite{Meskhidze:2024}] indicate that one \emph{must} generalize
the notion of torsion to recover a classical spacetime with torsion
from [teleparallel gravity]' (emphasis in original).  This is not
true, as the discussion in \cite{Schwartz:2023} shows: assuming that a
Lorentzian orthonormal tetrad converge to a classical tetrad according
to \eqref{eq:new_conv}, for any Lorentzian metric-compatible
derivative operator which has a regular Newtonian limit (in the sense
that its local connection form with respect to the classical tetrad
converges), one can show by direct computation that its torsion
converges to the torsion of the limiting derivative operator (this
generalises our above computation at the end of
\cref{sec:limit_assump}, which dealt with the special case of
derivative operators that parallelise the respective tetrads).  For
this conclusion, \emph{no consideration of extended coframes or the
  extended torsion is needed at all}, neither in its formulation nor
in its proof.  The mass torsion comes into play only when one starts
to consider either the gravitational field equations or the equations
of motion for matter, since these feature the \emph{contortion} of the
Lorentzian derivative operator.  In computing the contortion's
Newtonian limit, one needs to consider higher-order terms in the
post-Newtonian expansion of the Lorentzian orthonormal tetrad, which
precisely yield the extended coframe; as it turns out, the Newtonian
limit of the Lorentzian contortion depends not only on the Newtonian
limit of the torsion (i.e.\ the `usual' torsion of the limiting
derivative operator), but also on the mass torsion (see eq.\ (3.7) of
\cite{Schwartz:2023}).

Further, we want to comment on the role of spatial torsion in the
Newtonian limit of \acronym{TEGR} as discussed in ref.\
\cite{Schwartz:2023}.  In footnote 19 of ref.\ \cite{Meskhidze:2024},
it is stated that `[g]iven that the spatial torsion ultimately does
not seem to play any meaningful role in the theory, it is not clear
what the significance of this generalization to the theory [i.e.\
allowing for purely\footnote{We speak of `\emph{purely} spatial
  torsion' to emphasise that we mean only the purely spatial
  components of the torsion, i.e., in the notation of ref.\
  \cite{Meskhidze:2024}, $\tensor{T}{^a_{bc}} (f^i)_a (f_j)^b (f_k)^c$
  with $i,j,k \in \{2,3,4\}$.  The mixed spatio-temporal components
  $\tensor{T}{^a_{bc}} (f^i)_a (f_1)^b (f_j)^c$ with $i,j \in
  \{2,3,4\}$---which might be called part of the `spatial
  torsion'---can and do enter the field equations in ref.\
  \cite{Schwartz:2023}.  We also refer to the related discussion in
  section 3.3 of ref.\ \cite{March.etAl:2024}.} spatial torsion] is
supposed to be'.  To us, this statement seems rooted in a slight
misunderstanding of our intentions in article \cite{Schwartz:2023}.
In some sense, it is true that the purely spatial torsion does not
play a role when discussing the \emph{dynamical predictions} of
teleparallel Newton--Cartan gravity as presented in ref.\
\cite{Schwartz:2023}, since one may always consider an equivalent
solution of the gravitational field equations with vanishing purely
spatial torsion.  One of the main points of the discussion in ref.\
\cite{Schwartz:2023} is however the presented theory's
\emph{relationship to \acronym{TEGR}}, namely the former being the
latter's Newtonian limit.  When taking the Newtonian limit of a
solution (or, more precisely, of a one-parameter family of solutions)
of \acronym{TEGR} (assuming, of course, that it admits such a limit),
given in the form of a Lorentzian orthonormal tetrad and a compatible
flat derivative operator, in the general case the limiting derivative
operator \emph{will have non-vanishing purely spatial torsion}.  Thus,
to capture the Newtonian limit of teleparallel gravity in a precise
sense, including its geometric framework, one needs to allow for
non-vanishing purely spatial torsion.  The significance of purely
spatial torsion in teleparallel Newton--Cartan gravity as presented in
ref.\ \cite{Schwartz:2023} does not derive from any relevance to
dynamics---as correctly observed in ref.\ \cite{Meskhidze:2024}, it
has none; instead, it derives from the purely spatial torsion being a
necessary variable for parametrising the space of geometries arising
as the Newtonian limit of teleparallel gravity.

We also want to discuss an apparent misconception in ref.\
\cite{Meskhidze:2024} regarding the terminology `torsional
Newton--Cartan gravity'.  Concretely, this concerns ref.\
\cite{Meskhidze:2024}'s discussion of (a) recent work on the
coordinate-free formulation of the off-shell post-Newtonian expansion
of \acronym{GR} \cite{Van_den_Bleeken:2017,Hansen.EtAl:2020}, (b) the
discussion of the null reduction of \acronym{TEGR} in
ref.~\cite{Read.Teh:2018}, and (c)~the discussion (by one of the
present authors) of the Newtonian limit of \acronym{TEGR} in
ref.~\cite{Schwartz:2023}.  Regarding these works, the introduction of
ref.\ \cite{Meskhidze:2024} states the following:
\begin{quote}
  The claims above appear to either contradict the standard narrative
  or one another.  As presented, NCT [Newton--Cartan Theory] ought to
  be understood as the classical limit of GR.  But, as indicated in
  the first bullet [regarding ref.\ \cite{Hansen.EtAl:2020}] above,
  some claim that torsional Newton-Cartan geometry arises in the limit
  instead.  Read and Teh [ref.\ \cite{Read.Teh:2018}], meanwhile,
  claim that the limit of Teleparallel Gravity is standard Newtonian
  gravitation. Finally, disagreeing with both, Schwartz [ref.\
  \cite{Schwartz:2023}] claims that torsional Newton-Cartan geometry
  is not actually the limit of GR but, rather, it ought to arise in
  the limit of Teleparallel Gravity. What are we to make of these
  claims?
\end{quote}
This misrepresents the situation in the quoted literature---for
example, nowhere in ref.~\cite{Schwartz:2023} is the purported claim
made, neither is a disagreement with the results of the articles
\cite{Read.Teh:2018, Hansen.EtAl:2020} stated or implied.  This is
probably due to the following terminological misunderstanding.  The
name `torsional Newton--Cartan gravity' does \emph{not} designate just
any kind of generalised Newton--Cartan gravity with any arbitrary kind
of torsion.  Instead, `torsional Newton--Cartan gravity' is the
historically-grown name for generalisations of standard Newton--Cartan
gravity \emph{in which the clock one-form $t_a$ is not closed}, i.e.\
in which $\boldsymbol{\D}\boldsymbol{t} \ne 0$ (or, in index notation,
$\partial_{[a} t_{b]} \ne 0$).  The torsion of any derivative operator
compatible with $t_a$ necessarily satisfies $t_a \tensor{T}{^a_{bc}} =
(\boldsymbol{\D}\boldsymbol{t})_{bc} = 2 \partial_{[b} t_{c]}$, and
thus is non-vanishing in the case $\boldsymbol{\D}\boldsymbol{t} \ne
0$---hence the name.  However, also in the case of closed $t_a$, one
can consider derivative operators compatible with $t_a$ and $h^{ab}$
with non-vanishing torsion---in fact, choosing a unit timelike vector
field $v^a$, the spatially projected (with respect to $v^a$) torsion
$\tensor{T}{^a_{bc}} - v^a t_d \tensor{T}{^d_{bc}}$ may be arbitrarily
specified \cite{Bekaert.Morand:2016,Geracie.EtAl:2015,Schwartz:2024}.
Of course, such a situation might be called `Newton--Cartan gravity
with torsion'; however, it is \emph{not} to be called `torsional
Newton--Cartan gravity', which is specifically the established name
for situations with non-vanishing \emph{timelike} torsion
$\boldsymbol{t}(\boldsymbol{T}(\cdot,\cdot)) = \boldsymbol{\D}
\boldsymbol{t} \ne 0$.  (We also refer to the related discussion in
section 3.1 of ref.\ \cite{March.etAl:2024}.)  The resolution of the
purported contradictions in the literature claimed by ref.\
\cite{Meskhidze:2024} is then simply as follows.
\begin{itemize}[nosep]
\item The geometrically formulated Newtonian limit of \acronym{GR} in
  the standard sense (as, e.g., formulated by Malament
  \cite{Malament:1986}) is standard Newton--Cartan gravity with a
  torsion-free derivative operator.
\item To extend this customary picture and give an \emph{off-shell}
  formulation of geometric Newtonian gravity in terms of a
  \emph{variational principle}, understood as a limit of \acronym{GR},
  it is necessary \cite{Hansen.EtAl:2020} to consider torsional
  Newton--Cartan gravity, i.e.\ allow for a non-closed clock form.  As
  noted above, this implies that any compatible derivative operator
  necessarily has torsion.  In particular any such derivative operator
  cannot arise as the Newtonian limit of the Levi--Civita derivative
  operator of \acronym{GR} (on which also see below).  However, this
  is not claimed in ref.\ \cite{Hansen.EtAl:2020}: there, the
  (torsionful) compatible derivative operators are simply considered
  as helpful tools in the formulation of the theory, without ascribing
  any fundamental spatio-temporal meaning to them.
\item Ref.\ \cite{Read.Teh:2018} does not deal with the Newtonian
  limit of \acronym{TEGR}, but instead with its null reduction (i.e.\
  its quotient with respect to a lightlike symmetry), which is quite a
  different procedure.  Furthermore, that the resulting theory in
  ref.\ \cite{Read.Teh:2018} takes the form of standard Newtonian
  gravity is due to (a)~the assumption that $\boldsymbol{\D}
  \boldsymbol{t} = 0$, (b)~the assumption that the spatial torsion of
  the limiting derivative operator vanish\footnote{In ref.\
    \cite{Read.Teh:2018}, it is erroneously claimed that this follows
    automatically by the Bianchi identity.}, and (c)~a particular
  gauge fixing for the extended coframe.  Without additional gauge
  fixing, the null reduction of \acronym{TEGR} will probably give rise
  to a full teleparallel version of Newton--Cartan gravity, similar to
  that discussed in ref.\ \cite{Schwartz:2023}.
\item The discussion in ref.\ \cite{Schwartz:2023} shows what kind of
  Newtonian limit arises from \acronym{TEGR} under the assumption of
  closed clock form, $\boldsymbol{\D}\boldsymbol{t} = 0$.  The
  Newtonian limit of the Lorentzian metric-compatible derivative
  operator of \acronym{TEGR} is compatible with $t_a$ and $h^{ab}$,
  and hence due to $\boldsymbol{\D}\boldsymbol{t} = 0$ its timelike
  torsion vanishes.  However, as discussed above, in general it will
  have non-vanishing purely spatial torsion.  This shows that the
  (closed clock form) Newtonian limit of \acronym{TEGR} is a specific
  form of Newton--Cartan gravity \emph{with torsion}, but \emph{not}
  torsional Newton--Cartan gravity.
\end{itemize}
Thus, taking precise notice of what the actual statements of the
articles \cite{Hansen.EtAl:2020,Read.Teh:2018,Schwartz:2023} are, one
can reconcile all their purported contradictions.  In the cases where
the articles do in fact deal with a Newtonian limit (i.e.\ excluding
ref.\ \cite{Read.Teh:2018}), this reconciliation is \emph{not} due to
a recognition of any kind of `differing methodology' between
them---the alleged contradictions arise only when interpreting the
statements too imprecisely.

Finally, we want to comment on ref.\ \cite{Meskhidze:2024}'s
discussion of the first work \cite{Van_den_Bleeken:2017} regarding the
relation of torsional Newton--Cartan gravity (in the above-described
sense of a non-closed clock form) to \acronym{GR}.  In this work, it
is shown that by relaxing the condition that the Levi--Civita
derivative operator of \acronym{GR} converge in the Newtonian limit,
one arrives at a generalised version of Newton--Cartan gravity, which
naturally includes timelike torsion.  In ref.\ \cite{Meskhidze:2024}
it is claimed that ref.\ \cite{Van_den_Bleeken:2017} employs
`contradictory methodology', since its result is based on an
assumption---namely that the derivative operator need not necessarily
converge---that purportedly contradicts known results.  However, the
cited well-known result \cite{Malament:1986} does \emph{not} imply
that for any one-parameter family of Lorentzian metrics converging to
a pair of classical metric structures $t_a, h^{ab}$, the family of
associated Levi--Civita connections converges (which would, in fact,
contradict the assumption of ref.\ \cite{Van_den_Bleeken:2017}).
Instead, it \emph{only} reaches this conclusion \emph{if closedness of
  the limiting clock form $t_a$ is assumed}!  As soon as one allows
for a \emph{non}-closed clock form, as is explicitly done in ref.\
\cite{Van_den_Bleeken:2017}, the convergence result does no longer
hold, and one may consider families of metrics with \emph{diverging}
Levi--Civita connections.  Studying this novel situation is precisely
the subject of ref.\ \cite{Van_den_Bleeken:2017}, and not
contradictory in any sense.  As discussed above for ref.\
\cite{Hansen.EtAl:2020}, in ref.\ \cite{Van_den_Bleeken:2017} a
compatible \emph{torsionful} derivative operator is introduced in
order to conveniently formulate the field equations.  However, it is
never claimed in ref.\ \cite{Van_den_Bleeken:2017} that this
derivative would arise as the limit of the Lorentzian Levi--Civita
ones (which would be false).  Furthermore, the attempt in ref.\
\cite{Meskhidze:2024} to distinguish between two different ways of the
consideration of Newtonian limits, namely in terms of (a) the
convergence of families of models parametrised by a continuous
parameter $\lambda \to 0$ and (b) formal power series expansions in
the squared inverse of the speed of light $c^{-2}$, as possibly
leading to different results is misguided: when results in approach
(a) are proved, as is common, by assuming the limit in $\lambda$ to
have a `Taylor expansion' around $\lambda = 0$ to some order, this is
equivalent to proving the same result in approach (b) by considering a
formal expansion in $c^{-2}$ to the same order.  These two approaches
are really just two different ways of writing down the same arguments.

\section*{Acknowledgements}

We thank Domenico Giulini for valuable discussions.

\printbibliography

\end{document}
